\documentclass[12pt,a4paper]{article}

\usepackage{ifthen} 
\newboolean{pdflatex}
\setboolean{pdflatex}{true} 

\newboolean{articletitles}
\setboolean{articletitles}{true} 

\newboolean{uprightparticles}
\setboolean{uprightparticles}{false} 

\newboolean{inbibliography}
\setboolean{inbibliography}{false} 

\usepackage[top=1in, bottom=1.25in, left=1in, right=1in]{geometry}

\usepackage{microtype}
\usepackage{lineno}  
\usepackage{xspace} 
\usepackage{caption} 

\usepackage{graphicx}  
\usepackage{color}
\usepackage{subcaption}
\usepackage{colortbl}
\graphicspath{{./figs/}} 

\usepackage{amsmath} 
\usepackage{amssymb}
\usepackage{amsfonts}
\usepackage{upgreek} 

\newcommand*\patchAmsMathEnvironmentForLineno[1]{%
\expandafter\let\csname old#1\expandafter\endcsname\csname #1\endcsname
\expandafter\let\csname oldend#1\expandafter\endcsname\csname
end#1\endcsname
 \renewenvironment{#1}%
   {\linenomath\csname old#1\endcsname}%
   {\csname oldend#1\endcsname\endlinenomath}%
}
\newcommand*\patchBothAmsMathEnvironmentsForLineno[1]{%
  \patchAmsMathEnvironmentForLineno{#1}%
  \patchAmsMathEnvironmentForLineno{#1*}%
}
\AtBeginDocument{%
\patchBothAmsMathEnvironmentsForLineno{equation}%
\patchBothAmsMathEnvironmentsForLineno{align}%
\patchBothAmsMathEnvironmentsForLineno{flalign}%
\patchBothAmsMathEnvironmentsForLineno{alignat}%
\patchBothAmsMathEnvironmentsForLineno{gather}%
\patchBothAmsMathEnvironmentsForLineno{multline}%
\patchBothAmsMathEnvironmentsForLineno{eqnarray}%
}

\usepackage{hyperref}    
\usepackage[all]{hypcap} 

\usepackage{xspace} 
\usepackage{upgreek}

\def\lhcb {\mbox{LHCb}\xspace}

\def\babar  {\mbox{BaBar}\xspace}
\def\belle  {\mbox{Belle}\xspace}

\def\cdf    {\mbox{CDF}\xspace}

\def\MagUp {\mbox{\em Mag\kern -0.05em Up}\xspace}

\ifthenelse{\boolean{uprightparticles}}%
{

 \def\Pmu         {\ensuremath{\upmu}\xspace}                 
 \def\Pnu         {\ensuremath{\upnu}\xspace}                 
                  
 \def\Ppi         {\ensuremath{\uppi}\xspace}

 \def\PDelta      {\ensuremath{\Delta}\xspace}                 
 \def\PXi      {\ensuremath{\Xi}\xspace}                 
 \def\PLambda      {\ensuremath{\Lambda}\xspace}                 
 \def\PSigma      {\ensuremath{\Sigma}\xspace}                 
 \def\POmega      {\ensuremath{\Omega}\xspace}                 
 \def\PUpsilon      {\ensuremath{\Upsilon}\xspace}                 
 
 \def\PB      {\ensuremath{\mathrm{B}}\xspace}                 
                  
 \def\PD      {\ensuremath{\mathrm{D}}\xspace}

 \def\PK      {\ensuremath{\mathrm{K}}\xspace}

 \def\Pb      {\ensuremath{\mathrm{b}}\xspace}                 
 \def\Pc      {\ensuremath{\mathrm{c}}\xspace}

 \def\Pi      {\ensuremath{\mathrm{i}}\xspace}

 \def\Ps      {\ensuremath{\mathrm{s}}\xspace}

}
{

 \def\Pmu         {\ensuremath{\mu}\xspace}                 
 \def\Pnu         {\ensuremath{\nu}\xspace}                 
                  
 \def\Ppi         {\ensuremath{\pi}\xspace}

 \mathchardef\PDelta="7101
 \mathchardef\PXi="7104
 \mathchardef\PLambda="7103
 \mathchardef\PSigma="7106
 \mathchardef\POmega="710A
 \mathchardef\PUpsilon="7107
                  
 \def\PB      {\ensuremath{B}\xspace}                 
                  
 \def\PD      {\ensuremath{D}\xspace}

 \def\PK      {\ensuremath{K}\xspace}

 \def\Pb      {\ensuremath{b}\xspace}                 
 \def\Pc      {\ensuremath{c}\xspace}

 \def\Pi      {\ensuremath{i}\xspace}

 \def\Ps      {\ensuremath{s}\xspace}

}

\makeatletter
\ifcase \@ptsize \relax
  \newcommand{\miniscule}{\@setfontsize\miniscule{4}{5}}
\or
  \newcommand{\miniscule}{\@setfontsize\miniscule{5}{6}}
\or
  \newcommand{\miniscule}{\@setfontsize\miniscule{5}{6}}
\fi
\makeatother

\DeclareRobustCommand{\optbar}[1]{\shortstack{{\miniscule (\rule[.5ex]{1.25em}{.18mm})}
  \\ [-.7ex] $#1$}}

\def\mun        {{\ensuremath{\Pmu^-}}\xspace} 

\def\neub       {{\ensuremath{\overline{\Pnu}}}\xspace}

\def\neumb      {{\ensuremath{\neub_\mu}}\xspace}

\def\squark    {{\ensuremath{\Ps}}\xspace}

\def\cquark    {{\ensuremath{\Pc}}\xspace}

\def\bquark    {{\ensuremath{\Pb}}\xspace}

\def\pion   {{\ensuremath{\Ppi}}\xspace}

\def\pip    {{\ensuremath{\pion^+}}\xspace}
\def\pim    {{\ensuremath{\pion^-}}\xspace}

\def\kaon    {{\ensuremath{\PK}}\xspace}
  \def\Kbar    {{\kern 0.2em\overline{\kern -0.2em \PK}{}}\xspace}

\def\KorKbar    {\kern 0.18em\optbar{\kern -0.18em K}{}\xspace}

\def\Kzb     {{\ensuremath{\Kbar{}^0}}\xspace}
\def\Kp      {{\ensuremath{\kaon^+}}\xspace}
\def\Km      {{\ensuremath{\kaon^-}}\xspace}

\def\KS      {{\ensuremath{\kaon^0_{\mathrm{ \scriptscriptstyle S}}}}\xspace}

  \def\Dbar    {{\kern 0.2em\overline{\kern -0.2em \PD}{}}\xspace}
\def\D       {{\ensuremath{\PD}}\xspace}

\def\DorDbar    {\kern 0.18em\optbar{\kern -0.18em D}{}\xspace}
\def\Dz      {{\ensuremath{\D^0}}\xspace}
\def\Dzb     {{\ensuremath{\Dbar{}^0}}\xspace}
\def\Dp      {{\ensuremath{\D^+}}\xspace}
\def\Dm      {{\ensuremath{\D^-}}\xspace}

\def\Dstar   {{\ensuremath{\D^*}}\xspace}

\def\Dstarp  {{\ensuremath{\D^{*+}}}\xspace}
\def\Dstarm  {{\ensuremath{\D^{*-}}}\xspace}

\def\B       {{\ensuremath{\PB}}\xspace}
\def\Bbar    {{\ensuremath{\kern 0.18em\overline{\kern -0.18em \PB}{}}}\xspace}
\def\Bb      {{\ensuremath{\Bbar}}\xspace}
\def\BorBbar    {\kern 0.18em\optbar{\kern -0.18em B}{}\xspace}
\def\Bz      {{\ensuremath{\B^0}}\xspace}
\def\Bzb     {{\ensuremath{\Bbar{}^0}}\xspace}

\def\Bs      {{\ensuremath{\B^0_\squark}}\xspace}
\def\Bsb     {{\ensuremath{\Bbar{}^0_\squark}}\xspace}

  \def\Y#1S{\ensuremath{\PUpsilon{(#1S)}}\xspace}

\def\Lbar        {{\ensuremath{\kern 0.1em\overline{\kern -0.1em\PLambda}}}\xspace}
\def\LorLbar    {\kern 0.18em\optbar{\kern -0.18em \PLambda}{}\xspace}

\newcommand{\decay}[2]{\ensuremath{#1\!\to #2}\xspace}         
\def\ra                 {\ensuremath{\rightarrow}\xspace}
\def\to                 {\ensuremath{\rightarrow}\xspace}

\def\CP                {{\ensuremath{C\!P}}\xspace}

\def\AT#1     {\ensuremath{A_{\mathrm{T}}^{#1}}\xspace}           

\def\C#1      {\ensuremath{\mathcal{C}_{#1}}\xspace}                       
\def\Cp#1     {\ensuremath{\mathcal{C}_{#1}^{'}}\xspace}                    
\def\Ceff#1   {\ensuremath{\mathcal{C}_{#1}^{\mathrm{(eff)}}}\xspace}        
\def\Cpeff#1  {\ensuremath{\mathcal{C}_{#1}^{'\mathrm{(eff)}}}\xspace}       
\def\Ope#1    {\ensuremath{\mathcal{O}_{#1}}\xspace}                       
\def\Opep#1   {\ensuremath{\mathcal{O}_{#1}^{'}}\xspace}                    

\newcommand{\tev}{\ifthenelse{\boolean{inbibliography}}{\ensuremath{~T\kern -0.05em eV}\xspace}{\ensuremath{\mathrm{\,Te\kern -0.1em V}}}\xspace}
\newcommand{\gev}{\ensuremath{\mathrm{\,Ge\kern -0.1em V}}\xspace}
\newcommand{\mev}{\ensuremath{\mathrm{\,Me\kern -0.1em V}}\xspace}
\newcommand{\kev}{\ensuremath{\mathrm{\,ke\kern -0.1em V}}\xspace}
\newcommand{\ev}{\ensuremath{\mathrm{\,e\kern -0.1em V}}\xspace}
\newcommand{\gevc}{\ensuremath{{\mathrm{\,Ge\kern -0.1em V\!/}c}}\xspace}
\newcommand{\mevc}{\ensuremath{{\mathrm{\,Me\kern -0.1em V\!/}c}}\xspace}
\newcommand{\gevcc}{\ensuremath{{\mathrm{\,Ge\kern -0.1em V\!/}c^2}}\xspace}
\newcommand{\gevgevcccc}{\ensuremath{{\mathrm{\,Ge\kern -0.1em V^2\!/}c^4}}\xspace}
\newcommand{\mevcc}{\ensuremath{{\mathrm{\,Me\kern -0.1em V\!/}c^2}}\xspace}

\def\mum  {\ensuremath{{\,\upmu\mathrm{m}}}\xspace}

\def\invfb   {\ensuremath{\mbox{\,fb}^{-1}}\xspace}

\newcommand{\stat}{\ensuremath{\mathrm{\,(stat)}}\xspace}
\newcommand{\syst}{\ensuremath{\mathrm{\,(syst)}}\xspace}

\newcommand{\chisq}{\ensuremath{\chi^2}\xspace}

\newcommand{\chisqip}{\ensuremath{\chi^2_{\text{IP}}}\xspace}

\def\gsim{{~\raise.15em\hbox{$>$}\kern-.85em
          \lower.35em\hbox{$\sim$}~}\xspace}
\def\lsim{{~\raise.15em\hbox{$<$}\kern-.85em
          \lower.35em\hbox{$\sim$}~}\xspace}

\newcommand{\mean}[1]{\ensuremath{\left\langle #1 \right\rangle}} 

\def\sPlot{\mbox{\em sPlot}\xspace}

\def\pt         {\mbox{$p_{\mathrm{ T}}$}\xspace}

\def\tell1  {TELL1\xspace}
\def\ukl1   {UKL1\xspace}

\usepackage{cite} 
\usepackage{mciteplus}

\usepackage{longtable} 

\newcommand{\DACP}{\ensuremath{\Delta A_{\CP}}\xspace}
\newcommand{\KK}{\ensuremath{\Dz\to\Km\Kp}\xspace}
\newcommand{\PiPi}{\ensuremath{\Dz\to\pim\pip}\xspace}
\newcommand{\KPi}{\ensuremath{\Dz\to\Km\pip}\xspace}
\newcommand{\KPiPi}{\ensuremath{\Dp\to\Km\pip\pip}\xspace}

\newcommand{\KzbPi}{\ensuremath{\Dp\to\Kzb\pip}\xspace}
\newcommand{\deltam}{{\ensuremath{\delta m}}\xspace}

\newcommand{\acp}{\ensuremath{A_{\CP}}\xspace}
\newcommand{\acpkk}{\ensuremath{A_{\CP}(\Km\Kp)}\xspace}
\newcommand{\acppipi}{\ensuremath{A_{\CP}(\pim\pip)}\xspace}

\newcommand{\AcpKKFin}{0.14\pm0.15\stat\pm0.10\syst}

\newcommand{\AcpPiPi}{0.24}
\newcommand{\AcpPiPiStatErr}{0.15}
\newcommand{\AcpPiPiSysErr}{0.11}
\newcommand{\AcpLhcbPiPi}{0.07}
\newcommand{\AcpLhcbPiPiStatErr}{0.14}
\newcommand{\AcpLhcbPiPiSysErr}{0.11}
\newcommand{\AcpLHCb}{0.04}
\newcommand{\AcpLHCbStatErr}{0.12}
\newcommand{\AcpLHCbSysErr}{0.10}

\newcommand{\AcpSysErrPeaking}{0.015}

\newcommand{\KKSecFrac}{4.0}
\newcommand{\KPiSecFrac}{4.9}

\newcommand{\SecFracSys}{0.039}
\newcommand{\FittingSys}{0.025}

\begin{document}

\renewcommand{\thefootnote}{\fnsymbol{footnote}}
\setcounter{footnote}{1}

\begin{titlepage}
\pagenumbering{roman}

\vspace*{-1.5cm}
\centerline{\large EUROPEAN ORGANIZATION FOR NUCLEAR RESEARCH (CERN)}
\vspace*{1.5cm}
\noindent
\begin{tabular*}{\linewidth}{lc@{\extracolsep{\fill}}r@{\extracolsep{0pt}}}
\ifthenelse{\boolean{pdflatex}}
{\vspace*{-2.7cm}\mbox{\!\!\!\includegraphics[width=.14\textwidth]{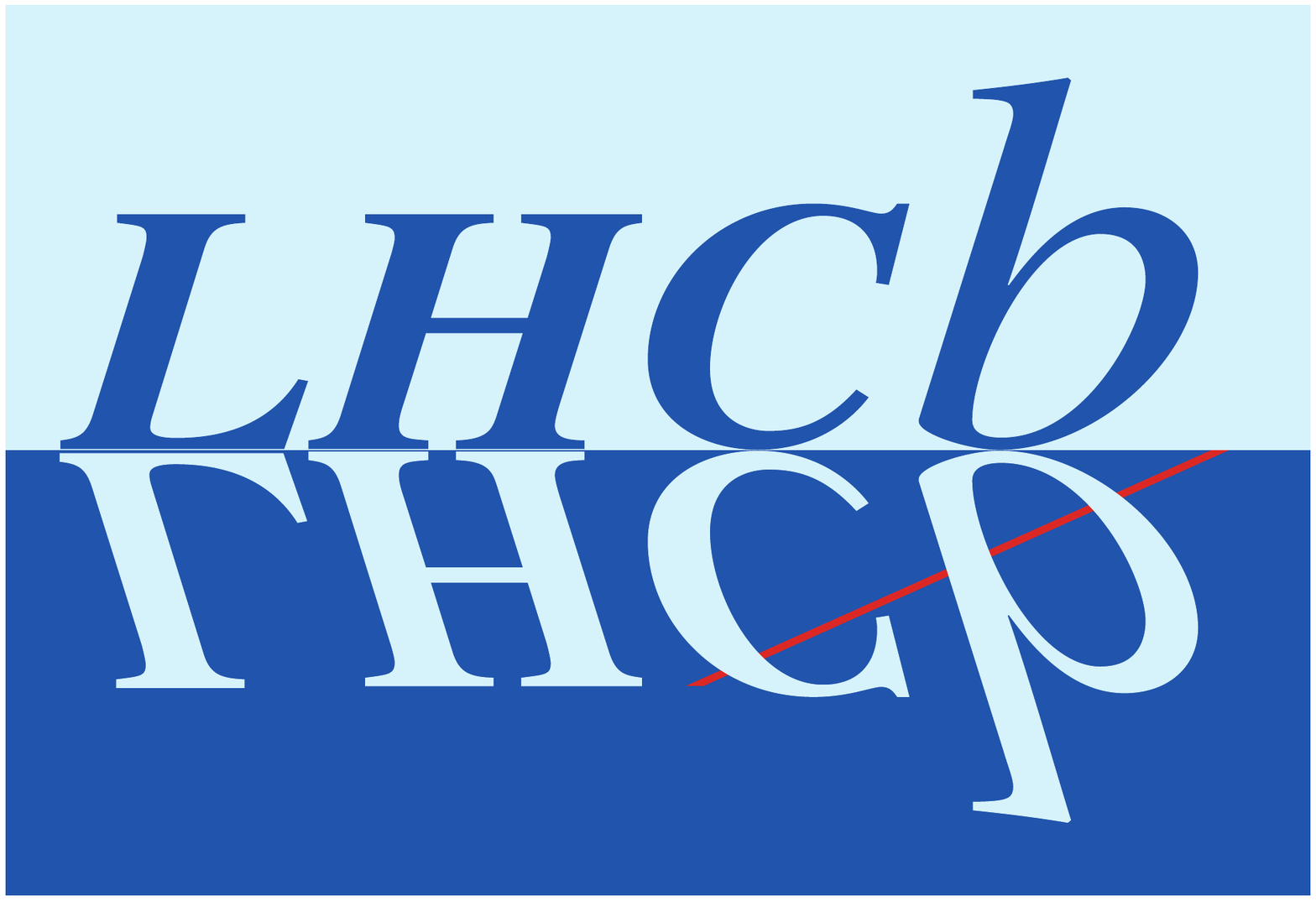}} & &}%
{\vspace*{-1.2cm}\mbox{\!\!\!\includegraphics[width=.12\textwidth]{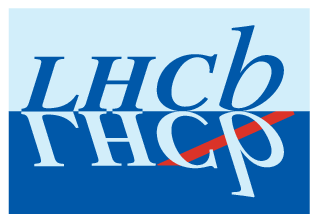}} & &}%
\\
 & & CERN-EP-2016-259 \\  
 & & LHCb-PAPER-2016-035 \\  
 & & October 29, 2016 \\ 
\end{tabular*}

\vspace*{1.0cm}

{\normalfont\bfseries\boldmath\huge
\begin{center}
Measurement of \CP asymmetry in \KK decays  \end{center}
}

\vspace*{0.5cm}

\begin{center}
The LHCb collaboration\footnote{Authors are listed at the end of this paper.}
\end{center}

\vspace{\fill}

\begin{abstract}
  \noindent
 
 A measurement of the time-integrated \CP asymmetry in the Cabibbo-suppressed decay \KK is performed using $pp$ collision data, corresponding to an integrated luminosity of 3\invfb, collected with the LHCb detector at centre-of-mass energies of 7 and 8\tev. The flavour of the charm meson at production is determined from the charge of the pion in
$\Dstarp\to\Dz\pip$ and $\Dstarm\to\Dzb\pim$ decays. The time-integrated \CP asymmetry \acpkk is obtained assuming negligible \CP violation in charm mixing and in Cabibbo-favoured \KPi, \KPiPi and \KzbPi decays used as calibration channels. It is found to be
\begin{align*}
\acpkk=(\AcpKKFin)\%.
\end{align*}
A combination of this result with previous LHCb measurements yields 
\begin{align*}
\acpkk=(\AcpLHCb\pm\AcpLHCbStatErr\stat \pm\AcpLHCbSysErr\syst)\%, \\
\acppipi=(\AcpLhcbPiPi\pm\AcpLhcbPiPiStatErr\stat \pm\AcpLhcbPiPiSysErr\syst)\%.
\end{align*}
These are the most precise measurements from a single experiment. The result for \acpkk is the most precise determination of a time-integrated \CP asymmetry in the charm sector to date, and neither measurement shows evidence of \CP asymmetry.
\end{abstract}

\vspace*{2.0cm}

\begin{center}
  Published in Phys.~Lett.~B
\end{center}

\vspace{\fill}

{\footnotesize 
\centerline{\copyright~CERN on behalf of the \lhcb collaboration, licence \href{http://creativecommons.org/licenses/by/4.0/}{CC-BY-4.0}.}}
\vspace*{2mm}

\end{titlepage}


\newpage
\setcounter{page}{2}
\mbox{~}

\cleardoublepage

\renewcommand{\thefootnote}{\arabic{footnote}}
\setcounter{footnote}{0}


\pagestyle{plain} 
\setcounter{page}{1}
\pagenumbering{arabic}

\section{Introduction}
\label{sec:Introduction}

In the Standard Model (SM), the violation of the charge-parity (\CP) symmetry is governed by an irreducible complex phase in the Cabibbo-Kobayashi-Maskawa (CKM) matrix.
Charmed hadrons provide the only way to probe \CP violation with up-type quarks.
Recent studies of \CP violation in weak decays of $D$ mesons have not shown evidence of \CP symmetry breaking~\cite{HFAG}, while its violation is well established in decays of mesons with down-type quarks (strange and beauty)
~\cite{Christenson:1964fg,Aubert:2001nu,Abe:2001xe,LHCb-PAPER-2013-018,LHCb-PAPER-2012-001}. 

The \CP-even decays\footnote{Throughout this Letter, charge conjugation is implicit unless otherwise stated.} \KK and \PiPi are singly Cabibbo-suppressed, and for these decays \Dz and \Dzb mesons share the same final state.
The amount of \CP violation in these decays is expected to be below the percent level~\cite{Feldmann:2012js, Bhattacharya:2012ah, Pirtskhalava:2011va,
Brod:2012ud, Cheng:2012xb, Muller:2015rna,Golden:1989qx,Li:2012cfa}, but large theoretical uncertainties due to long-distance interactions prevent precise SM predictions. In the presence of physics beyond the SM, the expected \CP asymmetries could be enhanced~\cite{Giudice:2012qq}, although an observation near the current experimental limits would be consistent with the SM expectation. The \CP asymmetries in these decays are sensitive to both direct and indirect \CP violation~\cite{HFAG, LHCb-PAPER-2015-055}.
The direct \CP violation is associated with the breaking of \CP symmetry in the decay amplitude.
Under $SU(3)$ flavour symmetry, the direct \CP asymmetries in the decays \KK and \PiPi are expected to have the same magnitudes and opposite sign~\cite{Grossman:2006jg}.
Indirect \CP violation, occurring through \Dz--\Dzb mixing and interference processes in the mixing and the decay, is expected to be small and is measured to be below $10^{-3}$~\cite{HFAG}.

The most recent measurements of the time-integrated individual \CP asymmetries in $\Dz\to\Km\Kp$ and $\Dz\to\pim\pip$ decays have been performed by the \lhcb~\cite{LHCb-PAPER-2014-013}, \cdf~\cite{CDF:2012qw}, \babar~\cite{bib:babarpaper2008} and \belle~\cite{bib:bellepaper2008,*Ko:2012jh} collaborations.

 The measurement in Ref.~\cite{LHCb-PAPER-2014-013} uses \Dz mesons produced in semileptonic \bquark-hadron decays ($\Bb\to\Dz\mun\neumb X$), where the charge of the muon is used to identify ({{tag}}) the flavour of the \Dz meson at production, while the other measurements use \Dz mesons produced in the decay of the $\Dstar(2010)^+$ meson, hereafter referred to as \Dstarp.
Charmed hadrons may be produced at the $pp$ collision point either directly, or in the instantaneous decays of excited charm states. These two sources are referred to as prompt. Charmed hadrons produced in the decays of \bquark-hadrons are called secondary charmed hadrons.

This Letter presents a measurement of the time-integrated \CP asymmetry in the \KK decay rates
\begin{align}
A_{\CP}(\Dz\to\Km\Kp)\equiv \frac{\Gamma(D^0\rightarrow K^-K^+)-\Gamma(\Dzb\rightarrow K^-K^+)}{\Gamma(D^0\rightarrow K^-K^+)+\Gamma(\Dzb\rightarrow K^-K^+)},
\end{align}
using a data sample of proton-proton ($pp$) collisions at centre-of-mass energies of 7 and 8\tev, collected by the LHCb detector in 2011 and 2012, corresponding to approximately 3\invfb of integrated luminosity.
To distinguish the two \CP-conjugate decays, the  flavour of the \Dz at production must be known.  In this analysis, the  flavour of the \Dz is tagged by the charge of the soft pion, $\pi^+_{s}$, in the strong decay $\Dstar^+\to\Dz\pi^+_{s}$. A combination with the recent measurement of the difference between the time-integrated \CP asymmetries of $D^0 \rightarrow K^-K^+$ and $D^0 \rightarrow \pi^-\pi^+$  decays, $\DACP \equiv \acpkk -\acppipi$, in prompt charm decays~\cite{LHCb-PAPER-2015-055} allows the determination of \acppipi taking into account the correlation  between \DACP and \acpkk. In addition, a combination of the measurements using prompt charm decays and the measurements using secondary charm decays from semileptonic \bquark-hadron decays~\cite{LHCb-PAPER-2014-013} at LHCb yields the most precise measurement of these quantities by a single experiment.

The method to determine \acpkk follows the strategy described in Ref. \cite{LHCb-PAPER-2014-013}. In the analysis of $D^{*+}\rightarrow D^0(\rightarrow K^-K^+)\pi_{s}^+$ decays, two nuisance asymmetries must be considered, the production asymmetry of the \Dstarp meson $A_P(D^{*+})$, and the detection asymmetry $A_D(\pi^+_{s})$ of the soft pion caused by non charge-symmetric interaction probabilities with the detector material and instrumental asymmetry. 
The measured raw asymmetry in the number of observed signal decays, defined as
\begin{align}
A_{\rm raw} \equiv \frac{N(\Dz\to\Km\Kp)-N(\Dzb\to\Km\Kp)}{N(\Dz\to\Km\Kp)+N(\Dzb\to\Km\Kp)},
\end{align}
is related to the \CP asymmetry via
\begin{align}
A_{\CP}(D^0\rightarrow K^-K^+)=A_{\rm raw}(D^0\rightarrow K^-K^+)-A_P(D^{*+})-A_D(\pi^+_{s}),
\label{eq:AsymsKK}
\end{align}
assuming that the asymmetries are small and that the reconstruction efficiencies can be factorised.
The decay $D^{*+}\rightarrow D^0(\rightarrow K^-\pi^+)\pi_{s}^+$ is used as a calibration channel to determine the production and detection asymmetries.
Since this decay is Cabibbo-favoured, a negligible \CP asymmetry is assumed. 
In contrast to the decay into two kaons, the final state $K^-\pi^+$ is not \CP symmetric.
Therefore, additional detection asymmetries arising from the final state particles are present, giving
\begin{align}
A_{\rm raw}(D^0\rightarrow K^-\pi^+)&=A_P(D^{*+})+A_D(\pi^+_{s})+A_D(K^-\pi^+).
\label{eq:AsymsKPi}
\end{align}
In order to evaluate the detection asymmetry of the final state $K^-\pi^+$, enhanced by the different interaction cross-sections of positively and negatively charged kaons in the detector material, the Cabibbo-favoured decay $D^+\rightarrow K^-\pi^+\pi^+$ is employed.
In analogy to the $\Dz\to\Km\pip$ decay, the raw asymmetry in this channel is given by
\begin{align}
A_{\rm raw}(D^+\rightarrow K^-\pi^+\pi^+)&=A_P(D^{+})+A_D(K^-\pi^+_{l})+A_D(\pi^+_{h}).
\label{eq:AsymsKPiPi}
\end{align} 
The pion with the lower transverse momentum, $\pi^+_{l}$, is chosen to cancel the effect of the detection asymmetry of the pion of the decay $D^0\rightarrow K^-\pi^+$. The remaining production asymmetry of the $D^+$ meson $A_P(D^{+})$, and the detection asymmetry of the other pion $\pi^+_{h}$ are eliminated by incorporating the Cabibbo-favoured decay $D^+\rightarrow \Kzb\pi^+$ in the measurement. There, the measured raw asymmetry consists of the production asymmetry $A_P(D^{+})$, the detection asymmetry of the neutral kaon $A_D(\Kzb)$, and the detection asymmetry of the pion $A_D(\pi^+)$
\begin{align}
A_{\rm raw}(D^+\rightarrow \Kzb\pi^+)&=A_P(D^{+})+A_D(\Kzb)+A_D(\pi^+).
\label{eq:AsymsKsPi}
\end{align}
The specific choice that the pion with the higher (lower) transverse momentum in the decay $D^+\rightarrow K^-\pi^+\pi^+$ is used to cancel the effect of the detection asymmetry of the pion in $D^+\rightarrow \Kzb\pi^+$ ($D^0\rightarrow K^-\pi^+$) is based on the comparison of the kinematic spectra of the respective pions.
The detection asymmetry $A_D(\Kzb)$ includes \CP violation, mixing and different cross-sections for the interaction of neutral kaons with the detector material.
However, all of these effects are known, and $A_D(\Kzb)$ is calculated to be small since only neutral kaons that decay within the first part of the detector are selected~\cite{LHCb-PAPER-2014-013}.
The combination of Eqs.~\ref{eq:AsymsKK}--\ref{eq:AsymsKsPi} yields an expression for $A_{\CP}(D^0\rightarrow K^-K^+)$ that only depends on measurable raw asymmetries and the calculable $\Kzb$ detection asymmetry,
\begin{eqnarray}
\label{eq:AsymComb}
A_{\CP}(D^0\rightarrow K^-K^+)&=&A_{\rm raw}(D^0\rightarrow K^-K^+)-A_{\rm raw}(D^0\rightarrow K^-\pi^+) \\ 
&+&A_{\rm raw}(D^+\rightarrow K^-\pi^+\pi^+)-A_{\rm raw}(D^+\rightarrow \Kzb\pi^+) \nonumber \\
&+&A_D(\Kzb). \nonumber
\end{eqnarray}

\section{Detector and event selection}
\label{sec:detector}

The LHCb detector~\cite{Alves:2008zz,LHCb-DP-2014-002} is a
single-arm forward spectrometer covering the pseudorapidity range $2 < \eta < 5$, designed for
the study of particles containing \bquark or \cquark quarks. 
The detector includes a high-precision tracking system
consisting of a silicon-strip vertex detector surrounding the $pp$
interaction region, a large-area silicon-strip detector located
upstream of a dipole magnet with a bending power of about
$4{\mathrm{\,Tm}}$, and three stations of silicon-strip detectors and straw
drift tubes placed downstream of the magnet.
The tracking system provides a measurement of the momentum of charged particles with
a relative uncertainty that varies from 0.5\% at low momentum to 1.0\% at 200\gevc. The minimum distance of a track to a primary vertex (PV), the impact parameter (IP), 
is measured with a resolution of $(15+29/\pt)\mum$,
where \pt is the component of the momentum transverse to the beam, in\,\gevc.

Different types of charged hadrons are distinguished using information
from two ring-imaging Cherenkov detectors. 
Photons, electrons and hadrons are identified by a calorimeter system consisting of
scintillating-pad and preshower detectors, an electromagnetic
calorimeter and a hadronic calorimeter. Muons are identified by a
system composed of alternating layers of iron and multiwire
proportional chambers.
The magnetic field inside the detector breaks the symmetry between trajectories of positively and negatively charged particles as the positive particles are deflected in one direction, and the negative particles in the opposite direction. Due to the imperfect symmetry of the detector, this can lead to detection asymmetries.
Periodically reversing the magnetic field polarity throughout data-taking almost cancels the effect.
The configuration with the magnetic field pointing upwards, MagUp (downwards, MagDown), bends positively (negatively)
charged particles in the horizontal plane towards the centre of the LHC ring.

The singly Cabibbo-suppressed decay mode \KK and the Cabibbo-favoured modes \KPi, \KPiPi and \KzbPi are selected, where the \Dz candidates come from the $\Dstarp\to\Dz\pip$ decay. The \Dstarp and \Dp candidates must satisfy an online event selection performed by a trigger, which consists of a hardware and software stage, and a subsequent offline selection. The hardware stage of the trigger is based on information from the calorimeter and muon systems, followed by a software stage, which applies a full event reconstruction. In order to avoid asymmetries arising from the hardware trigger, each of the four decay channels is required to satisfy a trigger that is independent of the decay considered.
Both the software trigger and offline event selection use kinematic variables and decay time to isolate the signal decays from the background. 
To ensure a cancellation of possible trigger asymmetries in the software stage, for each of the calibration channels a specification about which particle triggers the event is made.

All secondary particles from \Dz and \Dp decays are required to be significantly displaced from any primary $pp$ interaction vertex, and have momentum and transverse momentum \pt larger than a minimum value. 
The final state hadrons are combined into a \Dz (\Dp) candidate. The \Dstarp vertex is formed by $\Dz$ and $\pi_s^+$ candidates, and is constrained to coincide with the associated PV~\cite{Hulsbergen2005566}. Similarly, a vertex fit of the \Dp decay products is made, where the \Dp candidate is constrained to originate from the corresponding PV.
In \KzbPi decays, the neutral kaon is reconstructed via decays into two charged pions, which are dominated by decays of the short-lived neutral kaon, $\KS$.
The mass of the \Kzb meson is constrained to the nominal mass of the $\KS$ state~\cite{Olive:2016xmw}.
Decays of \decay{\KS}{\pip\pim} are reconstructed using only \KS mesons that decay early enough for the secondary pions to be reconstructed in the vertex detector.

Further requirements are placed on: the track fit quality;  the $\Dstarp$ and \Dz (\Dp) vertex fit quality;
 the \Dz (\Dp) meson transverse momentum and its decay distance; the smallest value of \chisqip, of both the \Dz (\Dp) candidate and its decay products with respect to all PVs in the event.
The \chisqip is defined as the difference between the vertex-fit $\chi^2$ of the PV reconstructed with and without the considered particle.
For \Dz candidates, a selection criterion is placed on the angle between the \Dz momentum in the laboratory frame and the momentum of the kaon or the pion in the \Dz rest frame.
For \Dp candidates, additional requirements on the pseudorapidities, momenta and transverse momenta of the particles are applied in order to match the kinematic distributions of the two \Dp decay modes.

Cross-feed backgrounds from \D meson decays with a
kaon misidentified as a pion, and vice versa, are reduced using particle identification requirements.
After these selection criteria, the dominant background in $\Dstarp \to \Dz\pi_s^+$ decays consists of genuine \Dz candidates paired with unrelated pions originating from the primary interaction vertex. The main background in the distributions of \KPiPi and \KzbPi decays is combinatorial.
For the \Dz channels, fiducial requirements are imposed to exclude kinematic regions having a large asymmetry in the soft pion reconstruction efficiency~\cite{LHCb-PAPER-2015-055}.
These regions occur because low momentum particles of one charge at large or small angles in the horizontal plane may be deflected either out of the detector acceptance or into the non-instrumented beam pipe region, whereas particles with the other charge are more likely to remain within the acceptance.
About 70\% of the selected candidates are retained after these fiducial requirements.

The \Dz candidates satisfying the selection criteria are accepted for further analysis if the mass difference $\delta m \equiv  m(h^+h^-\pi^+_s) -m(h^+h^-) $  for $h=K,\pi$ is in the range 139.77--151.57\mevcc. 
To reduce the combinatorial background, the mass of the reconstructed \Dz candidate is required to lie in the range 1850--1884\mevcc and 1847--1887\mevcc for \KK and \KPi decays, respectively. This window corresponds to about two standard deviations of the mass resolution, as estimated from a 
fit to the mass distribution of the charm meson candidates.
The \Dp candidates are selected by requiring the reconstructed mass to lie in a 1820--1920\mevcc mass window.

The data sample includes events with multiple \Dstarp candidates.
The majority of these events contains the same reconstructed \Dz meson combined with different soft pion candidates. The fraction of events with multiple candidates in the considered range of \deltam is about 6.5\% for \KK events and about 4.9\% for \KPi events. These fractions are approximately the same for each magnet polarity. One of the multiple candidates is randomly selected and retained, the others are discarded. 

The full data sets recorded in 2011 and 2012 at 7 and 8\tev, respectively, are used for this analysis. They correspond to an integrated luminosity of about 1\invfb and 2\invfb, respectively. In 2011, approximately 60\% of the data was recorded with magnet polarity MagDown, whereas in 2012 approximately the same amount of data was taken with each magnet polarity. The data are split into four subsamples according to the magnet polarity and the data-taking year. 

\section{Measurement of the asymmetries}
\label{sec:asymmetries}

The raw asymmetries and the signal yields are determined from binned likelihood fits to the $\delta m$ distributions in the \Dz decay modes, and to the invariant mass distributions $m(\Dp)$ in the \Dp channels.
The fits are simultaneous for both flavours and the background yields are allowed to differ between them.
The fits to the four decay channels are made independently in the four subsamples.

The signal shape of the \deltam distribution is described by the sum of three Gaussian functions, two of which have a common mean. 
The means and widths of the Gaussian distributions are allowed to differ between \Dz and \Dzb because of a possible charge-dependent bias in the measurement of the momentum, while all the other parameters are shared.
The background is described by
an empirical function consisting of the product of an exponential function and a power-law function modelling the phase-space threshold~\cite{LHCb-PAPER-2015-030}
\begin{equation}
\label{eq:bkg}
\mathcal{P}_{\rm bkg}(\delta m|A,B,\delta m_0)\propto \left( \delta m-\delta m_0 \right) ^A e^{- B(\delta m-\delta m_0)},
\end{equation}
where the threshold $\delta m_0$ is fixed to the known \pip mass~\cite{Olive:2016xmw}. The parameters $A$ and $B$ describe the shape and are common to \Dz and \Dzb decays.

The signal shape of the \Dp decays is described by the sum of two Gaussian distributions and a bifurcated Gaussian distribution. The bifurcated Gaussian distribution describes the asymmetric tails of the invariant mass distribution arising from radiative processes in the decay. The background is modelled by a single exponential function, with the same slope for the \Dp and \Dm states.

The production and detection asymmetries depend on the kinematics of the particles involved.
If the kinematic distributions are very different, this may lead to 
an imperfect cancellation of the nuisance asymmetries in \acpkk. 
To remove any residual effect, the kinematic distributions of the four decay channels are equalised by means of a weighting procedure~\cite{CDF:2012qw}. 
Fiducial regions where this weighting procedure is not possible due to a lack of events in one of the channels are already excluded by the requirements on kinematic variables of the D+ decays.
Fits to the $\deltam$ and $m(\Dp)$ distributions of the unweighted data samples are used to obtain the kinematic distribution of the signal component by disentangling the signal and background components with the \sPlot technique~\cite{Pivk:2004ty}.
Then, the normalised signal distributions of the four channels are compared.
To obtain the greatest possible statistical sensitivity, especially for the channel with the lowest yield $D^+\rightarrow \Kzb\pi^+$, the following order of the weighting steps is chosen: first, the \KPiPi kinematic distributions are weighted to reproduce the \KzbPi kinematics; second, \KPi distributions are weighted to reproduce the \KPiPi kinematics, and, last, the \KK distributions are weighted to reproduce the \KPi kinematics.
At each step, the weights already calculated in the previous steps are applied.
Some of the steps are repeated until a satisfactory agreement of the distributions is achieved.
The underlying \Dstarp kinematic distributions are independent of the \Dz decay mode, but the selection requirements can introduce differences for the $K^-K^+$ and $K^-\pi^+$ final states, which are observed in the kinematical distributions of the \Dstarp candidates.
The variables used for the weighting procedure are: \pt, $\eta$ and azimuthal angle $\varphi$ of the \Dstarp candidates; \pt and $\eta$ of the \Dp mesons; \pt, $\eta$ and $\varphi$ of the pion in the \KzbPi channel and of the higher-transverse-momentum pion in \KPiPi decays; $\pt, \eta$ and $\varphi$ of the kaon and the pion in the \KPi and \KPiPi modes. For the weighting of the \KPiPi decay to agree with the \KPi decay, the pion with the lower transverse momentum in the \KPiPi channel is used.
For all weighting steps, by default, each variable is divided in $20$ uniform bins. If necessary, the transverse momenta are transformed to the interval [$0$,$1$] to account for long tails in the distributions.
The procedure leads to a few events in scarcely populated bins having very large weights.
In order to mitigate such an effect, an upper bound to the weights is applied.

After applying the weights, the effective sample size is given by $N_{\rm eff} = ( \sum_{i=1}^{N} w_i)^2/( \sum_{i=1}^{N} w_i^2)$, where $w_i$ is the weight of candidate $i$ and $N$ is the total number of candidates. The numbers of signal decays determined from fits to the samples before and after weighting are given in Table~\ref{tab:yields}.

\begin{table}[pt]
\caption{Signal yields of the four channels before and after the kinematic weighting. In the case of the weighted samples, effective yields are given.}
\centering
\begin{tabular}{l| c c c c|c}
 \textbf{Channel} &\textbf{Before weighting}& \textbf{After weighting}  \\
\hline
$\KK$  & $5.56\,\text{M}$ & $ ~~1.63\,\text{M}$ \\
$\KPi$ & $32.4\,\text{M}$ & $ ~~2.61\,\text{M}$ \\
$\KPiPi$   & $37.5\,\text{M}$ &$13.67\,\text{M}$ \\
$\KzbPi$ & $1.06\,\text{M}$ & $ ~~1.06\,\text{M}$ \\
\end{tabular}
\label{tab:yields}
\end{table}

The detection asymmetry $A_D(\Kzb)$ of the neutral kaon is identified as one of the sources of the residual asymmetry. The method of calculation is described in full detail in Ref.~\cite{LHCb-PAPER-2014-013} and is applied here in the same way. Based on the reconstructed trajectories and a model of the detector material, the expected asymmetries are determined for all neutral kaon candidates individually and then averaged. The calculated values are $(-0.052\pm0.013)\%$ for 2011, and $(-0.054\pm0.014)\%$ for 2012 data.
The individual values for the different categories do not differ between samples taken with different magnet polarities.

\begin{figure}[pb]

\vspace{0em}

\centering
\begin{subfigure}{0.49\textwidth}
\includegraphics[scale=0.35,  trim = 0mm 0mm 0mm 0mm, clip]{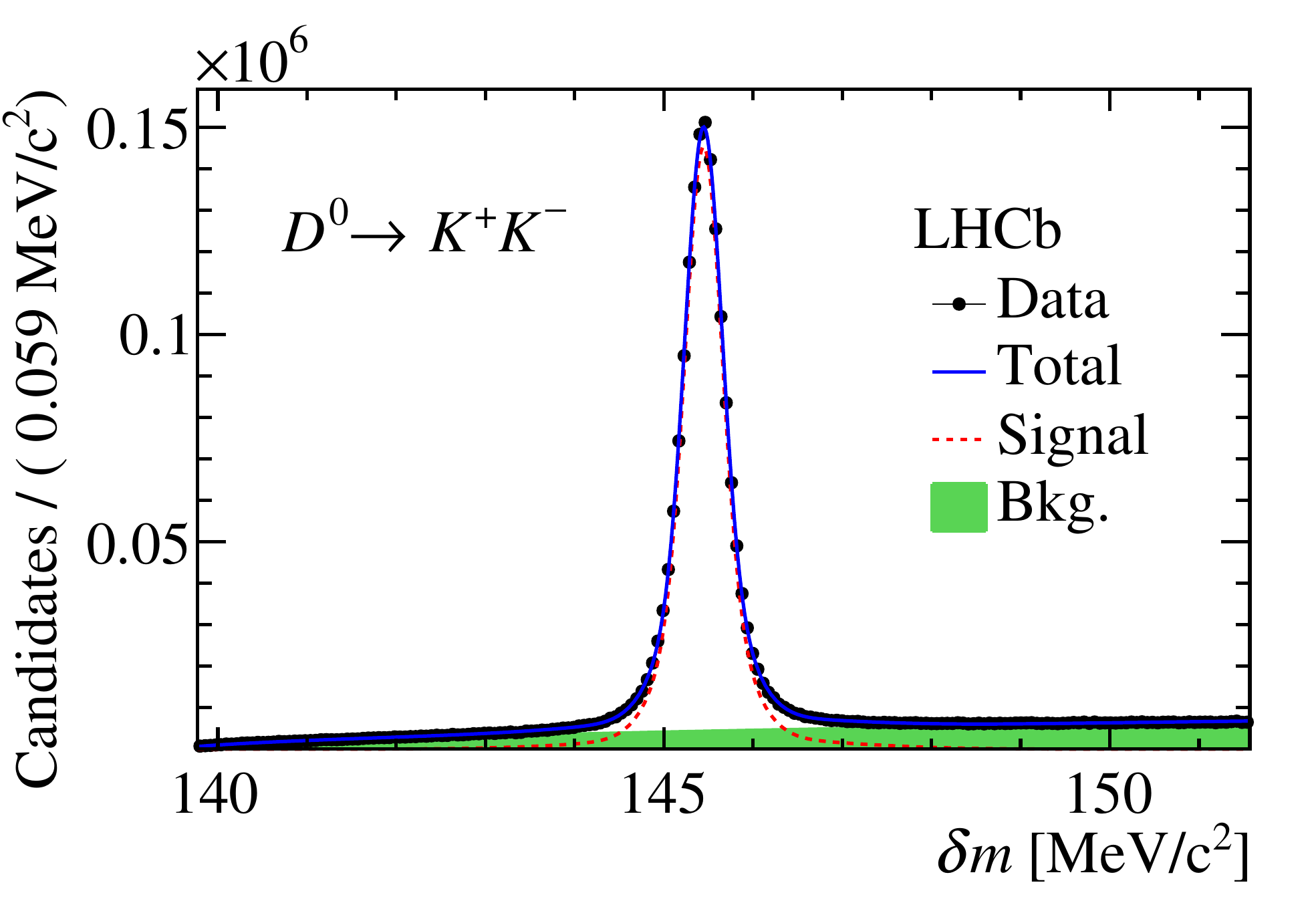}
\end{subfigure}
\begin{subfigure}{0.49\textwidth}
\includegraphics[scale=0.35,  trim = 0mm 0mm 0mm 0mm, clip]{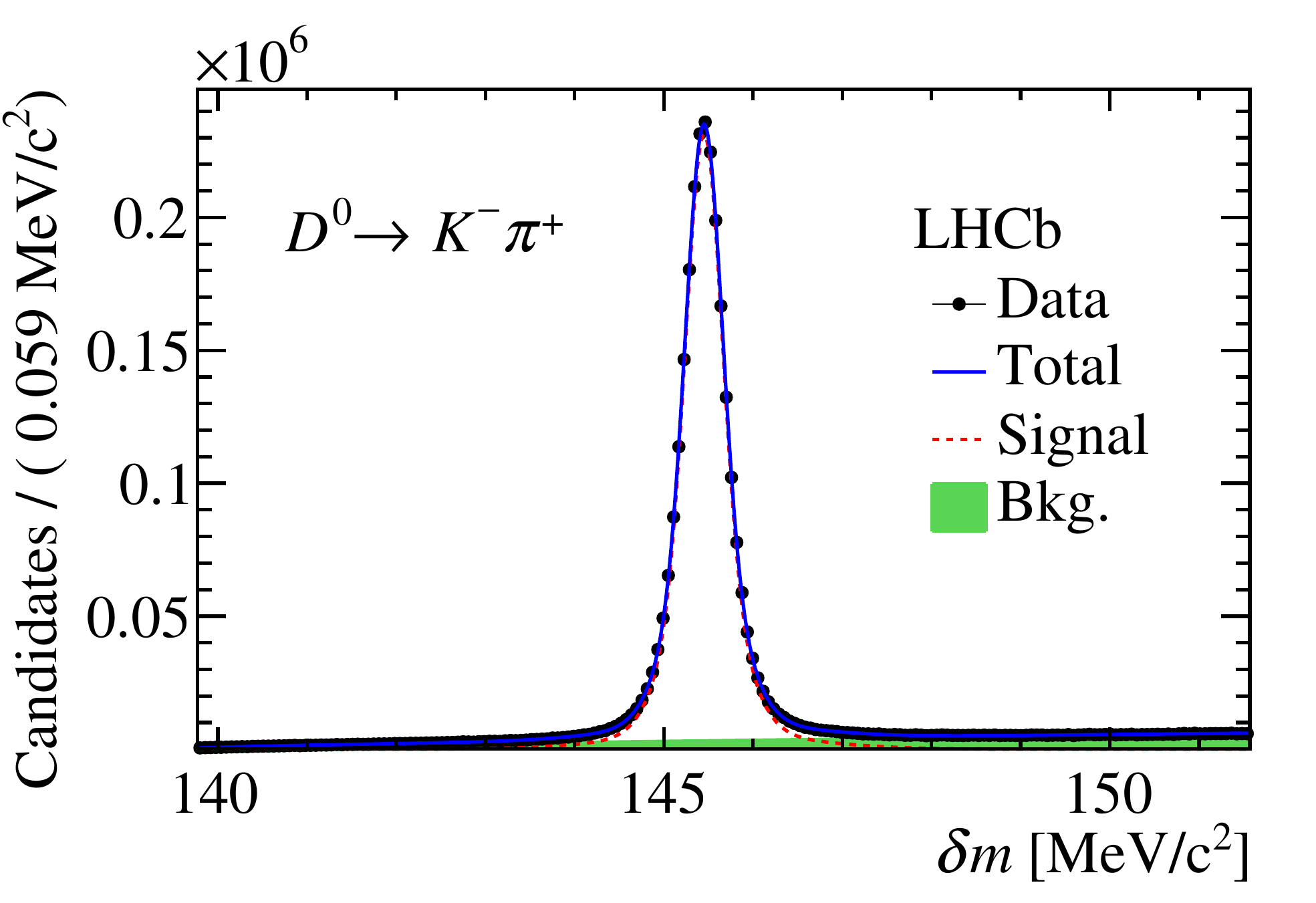}
\end{subfigure}
\begin{subfigure}{0.49\textwidth}
\includegraphics[scale=0.35,  trim = 0mm 0mm 0mm 0mm, clip]{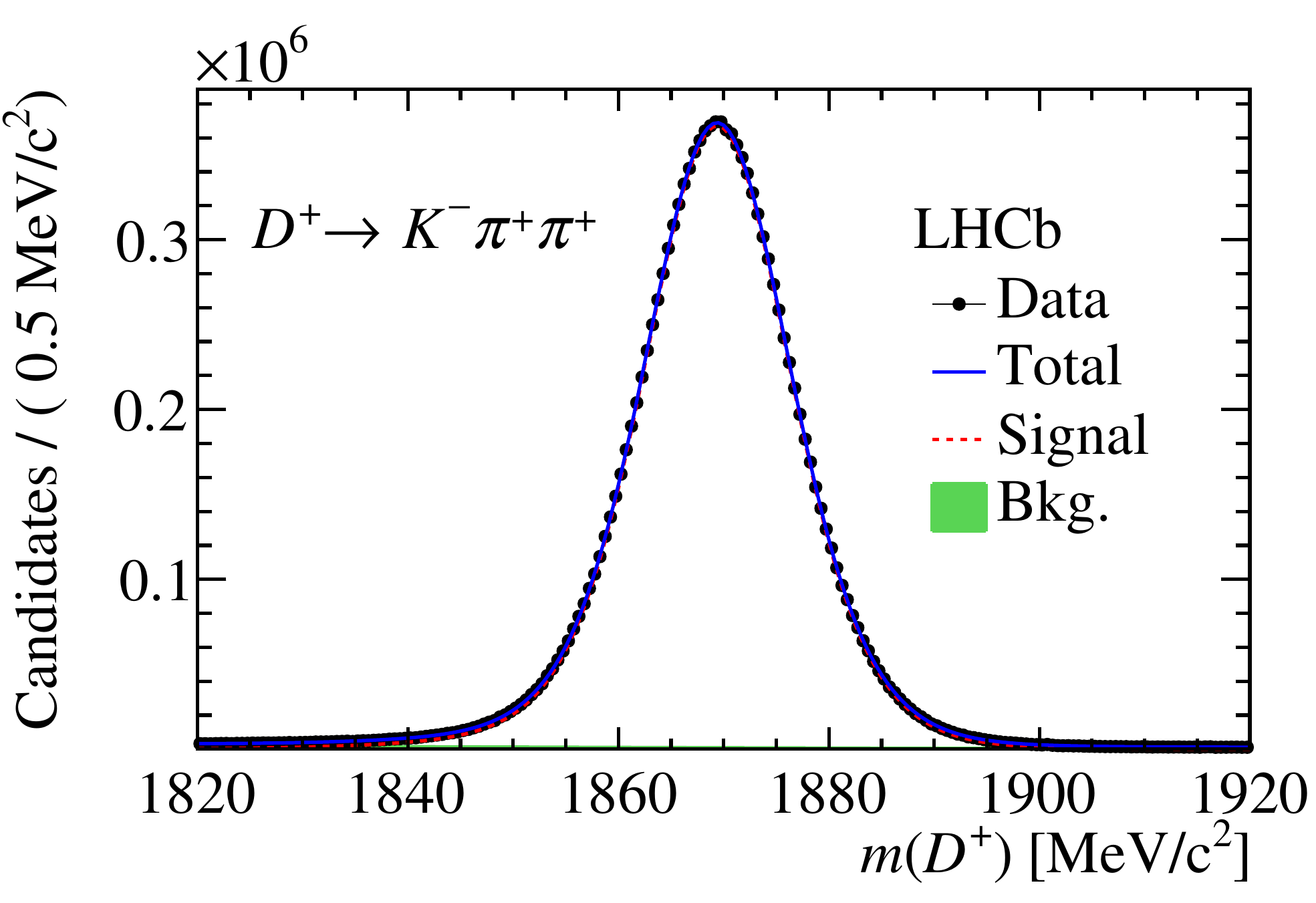}
\end{subfigure}
\begin{subfigure}{0.49\textwidth}
\includegraphics[scale=0.35,  trim = 0mm 0mm 0mm 0mm, clip]{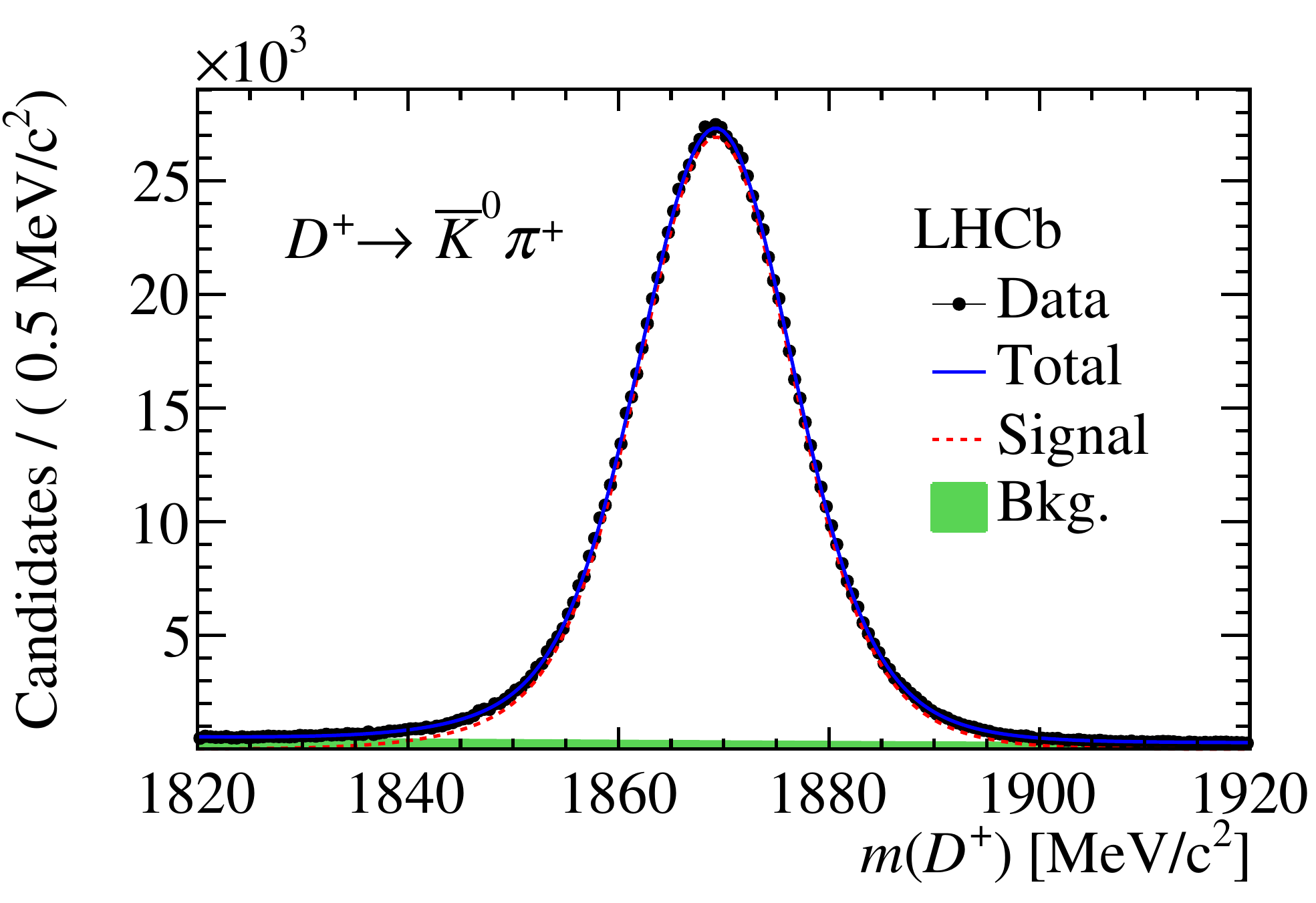}
\end{subfigure}
\caption{Fits to the $\deltam$ and to the $m(\Dp)$ distributions corresponding to the whole data sample and both flavours. Data samples after the kinematic weighting described in the text are used.}
\label{fig:fits}
\end{figure}

The raw asymmetries of the weighted samples, determined by the fits to the $\deltam$ and $m(\Dp)$ distributions shown in Fig.~\ref{fig:fits}, are presented in Table~\ref{tab:weightedAsyms}. The raw asymmetries are combined with the calculated detection asymmetry of the neutral kaon.
Testing the four independent measurements of \acpkk for mutual consistency gives $\chisq/\rm{ndf} = 0.80$, corresponding to a $p$-value of 0.50. The asymmetries obtained with the two magnet polarities within each year are arithmetically averaged
in order to ensure the cancellation of detection asymmetries which reverse sign with magnet polarity.
 The final result is then calculated as the weighted mean of the two data-taking periods. The weighted average of the values corresponding to all subsamples is calculated as $\acpkk = (0.14 \pm 0.15) \%$, where the uncertainty is statistical. The weighting procedure shifts the observed value of $\acpkk$ by $0.04\%$.

\begin{table}[tpb]

\vspace{-3em}

\centering
\caption{Measured asymmetries in $\%$ with their statistical uncertainties.}

\begin{tabular}{l| >{\centering\arraybackslash}m{3cm} >{\centering\arraybackslash}m{3cm} >{\centering\arraybackslash}m{3cm}}
\boldmath $2011$ & \textbf{MagUp} & \textbf{MagDown} & \textbf{Mean} \\
\hline
$A_{\rm raw}(D^0\rightarrow K^+K^-)$ & $-1.85\pm0.24$ & $\hphantom{-}0.05\pm0.20$ & $-0.90\pm0.16$ \\
$A_{\rm raw}(D^0\rightarrow K^-\pi^+)$ & $-2.87\pm0.18$ & $-1.43\pm0.15$ & $-2.15\pm0.12$ \\
$A_{\rm raw}(D^+\rightarrow K^-\pi^+\pi^+)$ & $-1.946\pm0.095$ & $-2.044\pm0.079$ & $-1.995\pm0.062$ \\
$A_{\rm raw}(\KzbPi)$ & $-0.95\pm0.30$ & $-0.93\pm0.25$ & $-0.94\pm0.20$ \\
$A_{D}(\Kzb)$ & $-0.052$ & $-0.052$ & $-0.052$ \\
\hline
$\acpkk$ & $-0.03\pm0.43$ & $\hphantom{-}0.32\pm0.37$ & $\hphantom{-}0.14\pm0.28$ 
\end{tabular}

\begin{tabular}{l| >{\centering\arraybackslash}m{3cm} >{\centering\arraybackslash}m{3cm} >{\centering\arraybackslash}m{3cm}}
\boldmath $2012\hphantom{+2012}$ & \textbf{MagUp} & \textbf{MagDown} & \textbf{Mean} \\
\hline
$A_{\rm raw}(\KK)$ & $-1.92\pm0.15$ & $-0.03\pm0.15$ & $-0.98\pm0.10$ \\
$A_{\rm raw}(\KPi)$ & $-2.23\pm0.11$ & $-1.65\pm0.11$ & $-1.939\pm0.079$ \\
$A_{\rm raw}(\KPiPi)$ & $-1.291\pm0.045$ & $-1.993\pm0.044$ & $-1.642\pm0.031$ \\
$A_{\rm raw}(\KzbPi)$ & $-0.92\pm0.17$ & $-0.83\pm0.17$ & $-0.88\pm0.12$ \\
$A_{D}(\Kzb)$ & $-0.054$ & $-0.054$ & $-0.054$ \\
\hline
$\acpkk$ & $-0.11\pm0.26$ & $\hphantom{-}0.40\pm0.26$ & $\hphantom{-}0.14\pm0.18$ 
\end{tabular}

\begin{tabular}{l| >{\centering\arraybackslash}m{3cm} >{\centering\arraybackslash}m{3cm} >{\centering\arraybackslash}m{3cm}}
\boldmath $2011+2012$ & \textbf{MagUp} & \textbf{MagDown} & \textbf{Mean} \\
\hline
$A_{\rm raw}(\KK)$ & $-1.90\pm0.12$ & $-0.01\pm0.12$ & $-0.95\pm0.10$ \\
$A_{\rm raw}(\KPi)$ & $-2.411\pm0.095$ & $-1.574\pm0.090$ & $-2.005\pm0.079$ \\
$A_{\rm raw}(\KPiPi)$ & $-1.411\pm0.041$ & $-2.005\pm0.038$ & $-1.714\pm0.031$ \\
$A_{\rm raw}(\KzbPi)$ & $-0.93\pm0.15$ & $-0.86\pm0.14$ & $-0.89\pm0.12$ \\
$A_{D}(\Kzb)$ & $-0.053$ & $-0.053$ & $-0.053$ \\
\hline
$\acpkk$ & $-0.09\pm0.22$ & $\hphantom{-}0.37\pm0.21$ & $\hphantom{-}0.14\pm0.15$ 
\end{tabular}

\label{tab:weightedAsyms}
\end{table}

\section{Systematic uncertainties}
\label{sec:systematics}
Possible systematic shifts of the measured \CP asymmetry can be caused by biases in the determination of individual raw asymmetries and non-cancellation of detection and production asymmetries. The determination of raw asymmetries is studied using several suitable alternative signal and background models in the fit of the mass distributions. Pseudoexperiments are generated based on the alternative fit results. The baseline model is fitted to the pseudoexperiment distributions. This is independently done for the four data categories and all channels.
The maximum observed deviations between the alternative results and the results of the fits to the generated pseudoexperiment distributions are combined, and a value of $\FittingSys\%$ is assigned as systematic uncertainty. This strategy allows systematic shifts and statistical fluctuations to be disentangled.

Partially reconstructed and misidentified three-body charm decays might produce a peaking background in the $\delta m$ distribution of the Cabibbo-suppressed decay \Dz\to\Km\Kp.
This background could therefore contribute to the signal yields obtained with the fit. If these incorrectly reconstructed decays were to have an asymmetry different from that of signal decays, the determined signal asymmetry would be shifted.
Simulated events are used to estimate the relative fraction of peaking background, which is then combined with production and detection asymmetries measured at LHCb~\cite{LHCb-PAPER-2014-013, LHCb-PAPER-2012-026} in order to obtain a conservative estimate of the asymmetry of this background.
A value of $\AcpSysErrPeaking\%$ is assigned as systematic uncertainty.

In order to test the influence of kinematic regions with high asymmetries of the final state particles in the channels \Dz\ra\Km\pip, \Dp\ra\Km\pip\pip and \Dp\ra\Kzb\pip, such regions are excluded in analogy to the treatment of the soft pion.
The difference in the values for $A_{\CP}(\Km\Kp)$, $0.040\%$, is taken as systematic uncertainty.

Possible incomplete cancellation of detection and production asymmetries is accounted for in several different ways. 
The weighting procedure is designed to equalise kinematic distributions, but a perfect agreement cannot be reached because of binning effects, the sequential rather than simultaneous weighting and the reduction of large weights. 
This effect is estimated by repeating the weighting with alternative configurations, which includes changing the number of bins, an alternative way of dealing with high weights and weighting in a reduced set of kinematic variables.
For each configuration, the \CP asymmetry is determined and the maximum deviation from the baseline result, $0.062\%$, is propagated as a systematic uncertainty.
Additionally, the kinematic dependence of the raw asymmetries observed in data are modelled with kinematically dependent detection and production asymmetries assigned to each particle in the decay.
These modelled detection and production asymmetries are then combined with the weighted kinematic distributions in data to calculate the raw asymmetries present in the individual channels.
The modelled raw asymmetries are combined to give the final \CP asymmetry according to Eq.~\ref{eq:AsymComb}.
Since no \CP asymmetry in \KK is included in this calculation, an ideal kinematic weighting, corresponding to a perfect cancellation of all detection and production asymmetries, would result in this \CP asymmetry being zero. The obtained deviation is 0.054\% and is treated as an independent systematic uncertainty.

Charmed mesons produced in the decay of beauty hadrons are suppressed by the requirement of a small \chisqip of the charm meson candidates with respect to the PV.
Nevertheless, a certain fraction of these decay chains passes the selection.
This leads to an effective production asymmetry that depends on the production asymmetry of the charm mesons, the production asymmetry of the beauty hadrons and the fraction of secondary charm decays.
The latter is determined for the \Dz decays by a fit to the \chisqip distributions when the selection requirement on this quantity is removed.
This yields an estimated secondary fraction of $\KKSecFrac\%$ for the channel \Dz\to\Km\Kp and $\KPiSecFrac\%$ for the channel \Dz\to\Km\pip.
For the \Dp decay channels, a conservative estimate of the difference in the fraction of secondary charm $f_{\rm sec}$ based on these numbers and on a comparison with simulated events, is made: $f_{\rm sec}(\Dp\to\Km\Kp\pip)-f_{\rm sec}(\Dp\to\Kzb\pip)=4.5\%$.
A combination of these numbers with the production asymmetries measured at LHCb \cite{LHCb-PAPER-2012-026, LHCb-PAPER-2014-053, LHCb-PAPER-2014-042, LHCb-PAPER-2015-032} yields a value of $\SecFracSys\%$, which is assigned as a systematic uncertainty.

The systematic uncertainty on the neutral kaon detection asymmetry is $0.014\%$.
Further sources of systematic uncertainty are investigated by performing consistency checks.
The analysis is repeated using more restrictive particle identification requirements and the result is found to be compatible with the baseline result.
Additionally, the measurement of the \CP asymmetry is repeated splitting the data-taking period into smaller intervals, and in bins of the momentum of the kaon in the decays \Dz\ra\Km\pip and \Dp\ra\Km\pip\pip.
No evidence of any dependence is found. All quoted systematic uncertainties are summarised in Table~\ref{tab:systs} and added in quadrature to obtain the overall systematic uncertainty.
\begin{table}
\caption{Systematic uncertainties from the different categories.
The quadratic sum is used to compute the total systematic uncertainty.
}
\label{tab:systs}
\centering
\begin{tabular}{l c}
\textbf{Category} & \textbf{Systematic uncertainty}\boldmath [$\%$]\\
\hline
Determination of raw asymmetries:\\

\hspace{1em}Fit model & $0.025$ \\

\hspace{1em}Peaking background & $0.015$ \\

Cancellation of nuisance asymmetries:\\
\hspace{1em}Additional fiducial cuts & $0.040$ \\

\hspace{1em}Weighting configuration & $0.062$ \\

\hspace{1em}Weighting simulation & $0.054$ \\

\hspace{1em}Secondary charm meson & $0.039$ \\

Neutral kaon asymmetry & $0.014$ \\
\hline 
\textbf{Total} & $0.10 $ 
\end{tabular}
\end{table}

\section{Summary and combination with previous LHCb measurements}
\label{sec:results}
The time-integrated \CP asymmetry in \KK decays is measured using data collected by the LHCb experiment and determined to be
\begin{align}
A_{\CP}(\Km\Kp)=(0.14\pm0.15\stat\pm0.10\syst)\%.
\end{align}
This result can be combined with previous LHCb measurements of the same and related observables. In Ref.~\cite{LHCb-PAPER-2014-013}, \acpkk was measured to be $A_{\CP}^{\rm sl}(\Km\Kp)=(-0.06\pm0.15\stat\pm0.10\syst)\%$ for \Dz mesons originating from semileptonic \bquark-hadron decays.
Since the same \Dp decay channels were employed for the cancellation of detection asymmetries, the result is partially correlated with the value presented in this Letter.
The statistical correlation coefficient is calculated as shown in Appendix~\ref{sec:correlations}, and is $\rho_{\rm stat}=0.36$ and the systematic uncertainties are conservatively assumed to be fully correlated.
A weighted average results in the following combined value for the \CP asymmetry in the \Dz\to\Km\Kp channel
\begin{align}
A_{\CP}^{\rm comb}(\Km\Kp)=(0.04\pm0.12\stat\pm0.10\syst)\%.
\end{align}
The difference in \CP asymmetries between \Dz\to\Km\Kp and \Dz\to\pim\pip decays, $\Delta A_{\CP}$, was measured at LHCb using prompt charm decays~\cite{LHCb-PAPER-2015-055}.
A combination of the measurement of $A_{\CP}(\Km\Kp)$ presented in this Letter with \DACP yields a value for $A_{\CP}(\pip\pim)$
\begin{align}
A_{\CP}(\pip\pim)=A_{\CP}(\Kp\Km)-\Delta A_{\CP}=(\AcpPiPi\pm\AcpPiPiStatErr\stat\pm\AcpPiPiSysErr\syst)\%.
\end{align}
The statistical correlation coefficient of the two measurements is $\rho_{\rm stat}=0.24$,
and the systematic uncertainties of the two analyses are assumed to be fully uncorrelated.

The correlation coefficient between this value and the measurement of $A_{\CP}^{\rm sl}(\pim\pip) = (-0.19\pm0.20\stat\pm0.10\syst)\%$ using semileptonically-tagged decays at LHCb~\cite{LHCb-PAPER-2014-013} is $\rho_{\rm stat}=0.28$. 
The weighted average of the values is
\begin{align*}
A_{\CP}^{\rm comb}(\pim\pip)=(\AcpLhcbPiPi\pm\AcpLhcbPiPiStatErr\stat\pm\AcpLhcbPiPiSysErr\syst)\%,
\end{align*}
where, again, the systematic uncertainties are assumed to be fully correlated. 
\begin{figure}[h]
\centering
\includegraphics[scale=0.5]{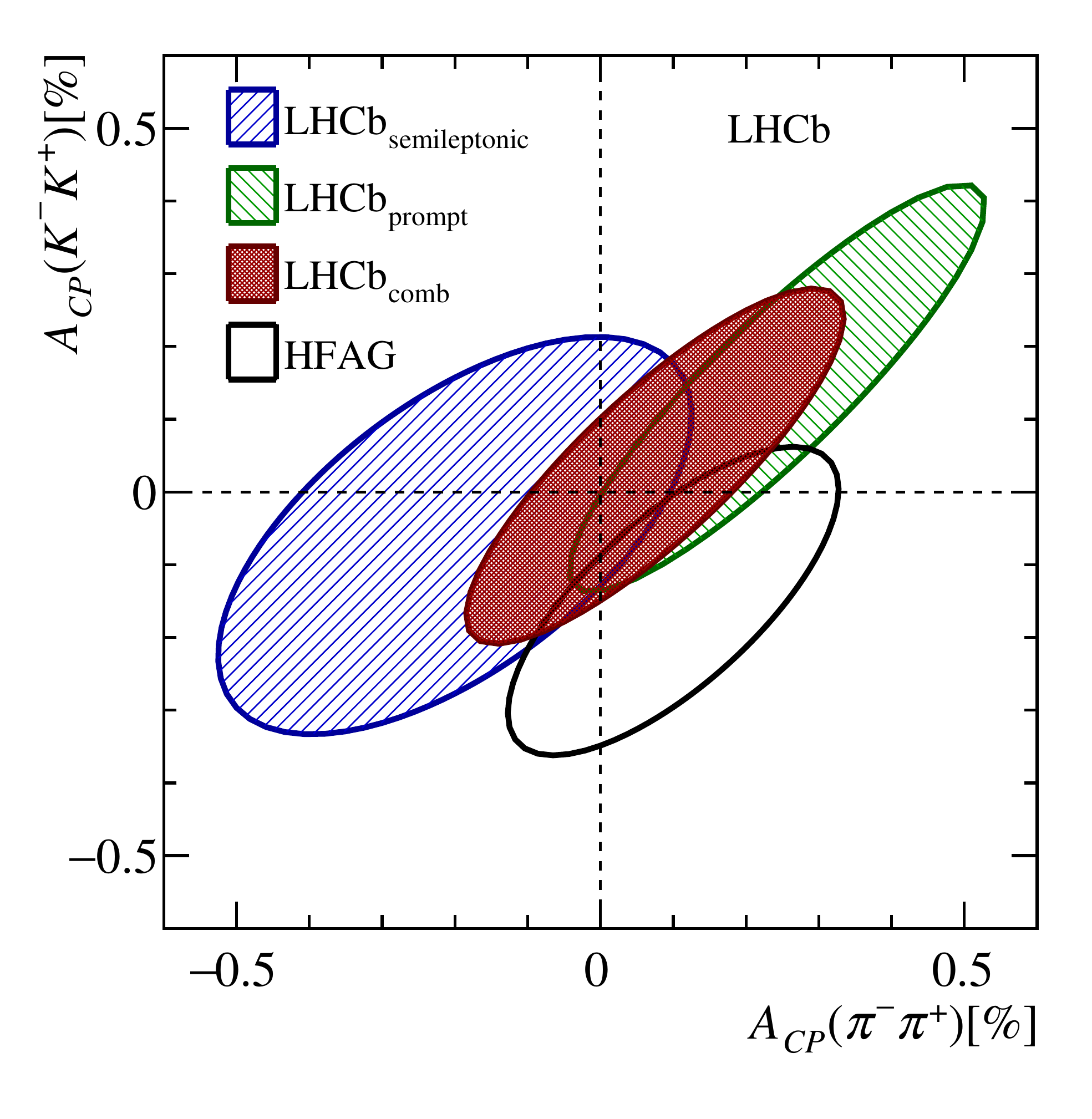}
\caption{Measurements of \CP violation asymmetries in \Dz\ra\Km\Kp and \Dz\ra\pim\pip decays.
Alongside the two LHCb measurements, presented in this Letter (green ellipse) and in Ref.\cite{LHCb-PAPER-2014-013} (blue ellipse), and their combination (red ellipse), the latest value of the Heavy Flavour Averaging Group \cite{HFAG} is shown (black ellipse). The latter already includes the measurement of \DACP with muon(pion)-tagged \Dz decays, using $3(1)\,$\invfb $pp$ collision data collected with the LHCb detector \cite{LHCb-PAPER-2014-013, LHCb-CONF-2013-003}. The $68\%$ confidence level contours are displayed where the statistical and systematic uncertainties are added in quadrature.}
\label{fig:AcpEllipses}
\end{figure}
When adding the statistical and systematic uncertainties in quadrature, the values for the \CP asymmetries in \Dz\to\Km\Kp and \Dz\to\pim\pip have a correlation coefficient $\rho_{\rm full}=0.61$.
Fig.~\ref{fig:AcpEllipses} shows the LHCb measurements of \CP asymmetry using both pion- and muon-tagged $\Dz\to\Km\Kp$ and $\Dz\to\pim\pip$ decays. Additionally, the latest combined values of the Heavy Flavour Averaging Group \cite{HFAG} for these quantities are presented.
The time-integrated \CP asymmetries can be interpreted in terms of direct and indirect \CP violation as shown in Appendix~\ref{sec:meandecaytimes}. 

In conclusion, no evidence of \CP violation is found in the Cabibbo-suppressed decays \KK and \PiPi. These results are obtained assuming that there is no \CP violation in \Dz--\Dzb mixing and no direct \CP violation in the Cabibbo-favoured \mbox{\KPi}, \KPiPi and \KzbPi decay modes. The combined LHCb results are the most precise measurements of the individual time-integrated \CP asymmetries \acpkk and \acppipi from a single experiment to date.

\section*{Acknowledgements}
\noindent We express our gratitude to our colleagues in the CERN
accelerator departments for the excellent performance of the LHC. We
thank the technical and administrative staff at the LHCb
institutes. We acknowledge support from CERN and from the national
agencies: CAPES, CNPq, FAPERJ and FINEP (Brazil); NSFC (China);
CNRS/IN2P3 (France); BMBF, DFG and MPG (Germany); INFN (Italy); 
FOM and NWO (The Netherlands); MNiSW and NCN (Poland); MEN/IFA (Romania); 
MinES and FASO (Russia); MinECo (Spain); SNSF and SER (Switzerland); 
NASU (Ukraine); STFC (United Kingdom); NSF (USA).
We acknowledge the computing resources that are provided by CERN, IN2P3 (France), KIT and DESY (Germany), INFN (Italy), SURF (The Netherlands), PIC (Spain), GridPP (United Kingdom), RRCKI and Yandex LLC (Russia), CSCS (Switzerland), IFIN-HH (Romania), CBPF (Brazil), PL-GRID (Poland) and OSC (USA). We are indebted to the communities behind the multiple open 
source software packages on which we depend.
Individual groups or members have received support from AvH Foundation (Germany),
EPLANET, Marie Sk\l{}odowska-Curie Actions and ERC (European Union), 
Conseil G\'{e}n\'{e}ral de Haute-Savoie, Labex ENIGMASS and OCEVU, 
R\'{e}gion Auvergne (France), RFBR and Yandex LLC (Russia), GVA, XuntaGal and GENCAT (Spain), Herchel Smith Fund, The Royal Society, Royal Commission for the Exhibition of 1851 and the Leverhulme Trust (United Kingdom).

\newpage
\clearpage

{\noindent\normalfont\bfseries\Large Appendices}

\appendix

\vspace*{1.0cm}

\section{Calculation of correlations}
\label{sec:correlations}

Since the measurement of $A_{\CP}(\Km\Kp)$ using semileptonic \bquark-hadron decays employs the same prompt $D^+$ calibration channels, it is correlated to the value obtained from prompt charm decays.
Due to different selection requirements and a different weighting procedure of the candidates, the asymmetries measured for the \Dp channels are not fully correlated.
The correlation factor $\rho$ between two weighted subsamples $X$ and $Y$ of a larger data sample $Z$ is given by
\begin{align}
\rho=\sqrt{\frac{\left(\sum_Z\omega_X\omega_Y\right)^2}{\sum_X\omega_X^2\sum_Y\omega_Y^2}},
\end{align}
where $\omega_X$ and $\omega_Y$ are the weights of candidates in the X and Y subsamples.
Whereas the four \mbox {\Dp\ra\KS\pip} data samples have correlation factors $\rho_{\KS\pi}$ between $0.64$ and $0.70$, the correlation factors of the \Dp\ra$\Km$\pip\pip samples, $\rho_{K\pi\pi}$, are in the range $0.07$ to $0.08$.
The main reason for these small correlations is the prescaling of the \Dp\ra$K$\pip\pip data in the semileptonic analysis which was removed in the prompt case.
From these numbers, for each data category the correlation $\rho_{A_{\CP}}$ of the values for $A_{\CP}(\Km\Kp)$ are calculated as
\begin{align}
\rho_{A_{\CP}}=\frac{1}{\sigma_{A_{\CP}}^{\mathrm {prompt}}\sigma_{A_{\CP}}^{\mathrm {sl}}}\left[\rho_{\KS\pi}\sigma_{\KS\pi}^{\mathrm {prompt}}\sigma_{\KS\pi}^{\mathrm {sl}}+\rho_{K\pi\pi}\sigma_{K\pi\pi}^{\mathrm {prompt}}\sigma_{K\pi\pi}^{\mathrm {sl}}\right]\label{eq:corrComb}.
\end{align}
Here, $\sigma$ represents the statistical uncertainty of the measured asymmetry of the respective channel.
This results in correlation factors between $0.34$ and $0.37$ for the four data categories.
When combining these correlations in a similar way to Eq.~\ref{eq:corrComb}, the statistical correlation $\rho_{\mathrm {stat}}$ between the semileptonic and prompt measurements of $A_{\CP}(\Kp\Km)$ is obtained to be $\rho_{\mathrm {stat}}=0.36$.

The other correlation factors presented in Sec.~\ref{sec:results} are obtained using a similar strategy.

\section{Mean decay times}
\label{sec:meandecaytimes}

The time-integrated \CP asymmetry $A_{\CP}(\Km\Kp)$ is not only sensitive to direct \CP violation, but also has a contribution from indirect \CP violation. This contribution depends on the mean decay time in units of the lifetime of the \Dz mesons, $\mean{t(hh)}/\tau(\Dz)$,
as
\begin{equation}
  A_{\CP} \approx a_{\CP}^{\rm dir} - A_{\Gamma} \frac{\mean{t(hh)}}{\tau(\Dz)} \ ,
 \label{eq:AcpDirectIndirect}
\end{equation}
where $a_{\CP}^{\rm dir}$ is the direct \CP violation term, $\tau$(\Dz) the \Dz
lifetime and $A_{\Gamma}$ a measure of indirect \CP violation.
More details about the method and the systematic uncertainties considered can be found in~\cite{LHCb-PAPER-2015-055}, \cite{LHCb-PAPER-2014-013}.

When calculating \acppipi from \acpkk and \DACP, a difference of the mean decay time of the \KK samples used for measuring \acpkk and \DACP leads to an additional contribution which is proportional to this difference and the size of indirect \CP violation. This can be accounted for by adding this difference to the mean decay time of the \PiPi sample used in the \DACP measurement. In Table~\ref{tab:averagetimes} this modified mean decay time is labelled by $\acpkk-\DACP$.

\begin{table}[ph]
\caption{Summary of the mean decay times of the $\Dz\to h^-h^+$ candidates used in the measurements of \DACP and \acp using prompt and semileptonic \Dz decays, and their combined values. The first uncertainty of the results is statistical, and the second one accounts for the systematics.}
\centering
\begin{tabular}{l|c|c|c|c}
\textbf {Tag}& \textbf {Mode} & \textbf {Measurement} & \textbf {$\mean{ t(hh)}/\tau(\Dz)$} & \textbf {Ref.} \\
\hline
Prompt &\Km\Kp &\DACP&$ 2.1524 \pm 0.0005 \pm0.0162 $&\cite{LHCb-PAPER-2015-055}\\
Prompt &\pim\pip &\DACP & $2.0371 \pm 0.0005 \pm 0.0151 $&\cite{LHCb-PAPER-2015-055}  \\
Prompt &\Km\Kp &\acpkk& $2.2390 \pm 0.0007 \pm 0.0187 $& --\\
Prompt &\pim\pip &\acpkk$-$\DACP &$2.1237 \pm 0.0008 \pm 0.0375 $& --\\
Semileptonic &\Km\Kp & \DACP & $1.082 \pm 0.001 \pm 0.004 $& \cite{LHCb-PAPER-2014-013} \\
Semileptonic &\pim\pip &\DACP & $ 1.068 \pm 0.001 \pm 0.004 $&\cite{LHCb-PAPER-2014-013}  \\
Semileptonic &\Km\Kp &\acpkk & $1.051 \pm 0.001 \pm 0.004 $&\cite{LHCb-PAPER-2014-013} \\
Semileptonic &\pim\pip &$\acpkk-\DACP$ &$1.0370 \pm 0.0011 \pm 0.0089 $& --\\
Pr. + sl. &\pim\pip &$\acpkk-\DACP$ & $1.7121 \pm 0.0007 \pm 0.0267 $& --\\
Pr. + sl.& \Km\Kp&\acpkk & $1.6111\pm 0.0007 \pm 0.0109 $ &--  \\
\end{tabular}
\label{tab:averagetimes}
\end{table}

\addcontentsline{toc}{section}{References}
\setboolean{inbibliography}{true}
\bibliographystyle{LHCb}
\providecommand{\href}[2]{#2}

\newpage
\centerline{\large\bf LHCb collaboration}
\begin{flushleft}
\small
R.~Aaij$^{40}$,
B.~Adeva$^{39}$,
M.~Adinolfi$^{48}$,
Z.~Ajaltouni$^{5}$,
S.~Akar$^{6}$,
J.~Albrecht$^{10}$,
F.~Alessio$^{40}$,
M.~Alexander$^{53}$,
S.~Ali$^{43}$,
G.~Alkhazov$^{31}$
% ...

{\footnotesize \it
% ...
$ ^{5}$Clermont Universit{\'e}, Universit{\'e} Blaise Pascal, CNRS/IN2P3, LPC, Clermont-Ferrand, France\\
$ ^{6}$CPPM, Aix-Marseille Universit{\'e}, CNRS/IN2P3, Marseille, France\\
% ...
$ ^{10}$Fakult{\"a}t Physik, Technische Universit{\"a}t Dortmund, Dortmund, Germany\\
% ...
$ ^{31}$Petersburg Nuclear Physics Institute (PNPI), Gatchina, Russia\\
% ...
$ ^{39}$Universidad de Santiago de Compostela, Santiago de Compostela, Spain\\
$ ^{40}$European Organization for Nuclear Research (CERN), Geneva, Switzerland\\
% ...
$ ^{43}$Nikhef National Institute for Subatomic Physics, Amsterdam, The Netherlands\\
% ...
$ ^{48}$H.H. Wills Physics Laboratory, University of Bristol, Bristol, United Kingdom\\
% ...
$ ^{53}$School of Physics and Astronomy, University of Glasgow, Glasgow, United Kingdom\\
% ...
}
\end{flushleft}

\end{document}